\documentclass[twocolumn,trackchanges]{aastex631}
\usepackage{subfiles} 
\usepackage{comment}
\usepackage{here}
\usepackage{amsmath}
\usepackage[T1]{fontenc}
\usepackage{scrextend}

\accepted{October 9, 2023}

\shorttitle{BH masses of high-z low-luminosity quasars}
\shortauthors{Takahashi et al.}
\graphicspath{{./}{./}}

\begin{document}

\title{Subaru High-$z$ Exploration of Low-luminosity Quasars (SHELLQs). XVII. Black Hole Mass Distribution at \boldmath{\ $z \sim 6$} Estimated via Spectral Comparison with Low-$z$ Quasars}

\author[0000-0003-3769-6630]{Ayumi Takahashi}
\affiliation{Graduate School of Science and Engineering, Ehime University, 2-5 Bunkyo-cho, Matsuyama, Ehime 790-8577, Japan;takahashi@cosmos.phys.sci.ehime-u.ac.jp}

\author[0000-0001-5063-0340]{Yoshiki Matsuoka}
\affiliation{Research Center for Space and Cosmic Evolution, Ehime University, Matsuyama, Ehime 790-8577, Japan}

\author[0000-0003-2984-6803]{Masafusa Onoue}
\affiliation{Kavli Institute for Astronomy and Astrophysics, Peking University, Beijing 100871, P.R.China}

\author[0000-0002-0106-7755]{Michael A. Strauss}
\affiliation{Department of Astrophysical Sciences, Princeton University, Peyton Hall, Princeton, NJ 08544, USA.}

\author[0000-0003-3954-4219]{Nobunari Kashikawa}
\affiliation{Department of Astronomy, School of Science, The University of Tokyo, Tokyo 113-0033, Japan}

\author[0000-0002-3531-7863]{Yoshiki Toba}
\affiliation{Research Center for Space and Cosmic Evolution, Ehime University, Matsuyama, Ehime 790-8577, Japan}
\affiliation{National Astronomical Observatory of Japan, 2-21-1 Osawa, Mitaka, Tokyo 181-8588, Japan}
\affiliation{Academia Sinica Institute of Astronomy and Astrophysics, 11F of Astronomy-Mathematics Building, AS/NTU, No.1, Section 4, Roosevelt Road, Taipei 10617, Taiwan}

\author[0000-0002-4923-3281]{Kazushi Iwasawa}
\affiliation{ICREA and Institut de Ci\`{e}ncies del Cosmos, Universitat de Barcelona, IEEC-UB, Mart'{i} i Franqu\`{e}s, 1, E-08028 Barcelona, Spain}

\author[0000-0001-6186-8792]{ Masatoshi Imanishi}
\affiliation{National Astronomical Observatory of Japan, Mitaka, Tokyo 181-8588, Japan}
\affiliation{Department of Astronomical Science, Graduate University for Advanced Studies (SOKENDAI), Mitaka, Tokyo 181-8588, Japan 8}

\author[0000-0002-2651-1701]{Masayuki Akiyama}
\affiliation{Astronomical Institute, Tohoku University, Aoba, Sendai, 980-8578, Japan}

\author[0000-0002-3866-9645]{Toshihiro Kawaguchi}
\affiliation{Department of Economics, Management and Information Science,
  Onomichi City University, Hisayamada 1600-2, Onomichi, Hiroshima 
722-8506, Japan}

\author[0000-0002-5197-8944]{Akatoki Noboriguchi}
\affiliation{Center for General Education, Shinshu University, 3-1-1 Asahi, Matsumoto, Nagano 390-8621, Japan}

\author[0000-0003-1700-5740]{Chien-Hsiu Lee}
\affiliation{ W. M. Keck Observatory, 65-1120 Mamalahoa Hwy, Kamuela, HI
96743, USA}




\begin{abstract}

We report the distribution of black hole (BH) masses and Eddingont ratios estimated for a sample of 131 low luminosity quasars in the early cosmic epoch ($5.6 < z < 7.0$). 
Our work is based on Subaru High-$z$ Exploration of Low-Luminosity Quasars (SHELLQs) project, which has constructed a low luminosity quasar sample down to 
$M_{1450} \sim \mathrm{- 21\ mag}$, exploiting the survey data of Hyper Suprime-Cam installed on Subaru Telescope. The discovery spectra of these quasars are limited to the rest-frame wavelengths of $\sim$ 1200 -- 1400 \AA, which contains no emission lines that can be used as  BH mass estimators. In order to overcome this problem, we made use of low-$z$ counterpart spectra from the Sloan Digital Sky Survey, which are spectrally matched to the high-$z$ spectra in overlapping wavelengths. We then combined the C~{\sc iv} emission line widths of the counterparts with the continuum luminosity from the SHELLQs data to estimate BH masses. The resulting BH mass distribution has a range of $\sim 10^{7-10} M_{\odot}$, with most of the quasars having BH masses $\sim 10^{8} M_{\odot}$ with sub-Eddington accretion. The present study provides not only a new insight into normal quasars in the reionization epoch, but also a new promising way to estimate BH masses of high-$z$ quasars without near-infrared spectroscopy.

\end{abstract}



\section{Introduction}
\label{intro}

Measuring BH masses of  high-$z$ ($z \gtrsim 6$) quasars is essential to understanding the evolution of supermassive black holes (SMBHs) and their host galaxies, as well as the mechanism of mass accretion and quasar radiation in the early universe. High-$z$ quasar surveys have been carried out over the past two decades with deep and wide-field observations, mainly in the optical to near-infrared (NIR) wavelengths. Now, the number of high-$z$ quasar samples has reached more than 300 \citep{2000AJ....120.1167F, 2003AJ....125.1649F, 2004AJ....128..515F, 2006AJ....131.1203F, 2008AJ....135.1057J,2009AJ....138..305J,2015AJ....149..188J, 2005ApJ...633..630W, 2007AJ....134.2435W, 2009AJ....137.3541W,
2010AJ....140..546W, 2010AJ....139..906W, 2014AJ....148...14B, 2016ApJS..227...11B, 2015ApJ...801L..11V, 2017ApJ...849...91M, 
2015MNRAS.454.3952R, 2017MNRAS.468.4702R, 2019MNRAS.487.1874R, 2019AJ....157..236Y, 2017ApJ...839...27W, 2018ApJ...869L...9W, 
2019ApJ...884...30W,2011Natur.474..616M, 2013ApJ...779...24V, 2022arXiv221206907F} including the most distant quasars known at $z > 7.5$  \citep[][]{2018Natur.553..473B, 2020ApJ...897L..14Y, wang2021luminous}. Despite the many quasars discovered in the early universe, it is unclear how such massive objects at the center of galaxies form and grow; the SMBHs in the most luminous high-$z$ quasars have masses of $M_{\rm BH} >  10^{9}\ M_{\odot}$ in a short time ($< 1$ Gyr) from Big Bang. For reference, the time needed for a BH with the mass of < $10^{5} M_{\odot}$ to build up a SMBH with $10^{9} M_\odot$ is at least 1 Gyr, assuming Eddington-limited accretion (see a recent review by Inayoshi, Visbal \& Haiman \citeyear{2020ARA&A..58...27I}).
No theoretical models that can fully explain this very early formation and growth have been established yet, but there are a few candidates attracting support
\citep[e.g.,][]{2012Sci...337..544V, haiman2013formation}. The first one assumes that the progenitors are the remnants of Pop-III stars with the mass of a few 10 --100 $\mathrm{M_{\odot}}$, which collapse into BH seeds and then grow very efficiently to $M_{\rm BH} \sim 10^5\ M_{\odot}$ \citep[e.g.][]{loeb1994collapse, oh2002second, bromm2003formation,
begelman2006formation, lodato2006supermassive, inayoshi2014formation, regan2014numerical, becerra2015formation, latif2016witnessing, volonteri2005rapid}. The second one is the dense star cluster scenario; a dense star cluster forms when locally unstable gas flows toward the galaxy center, then the stars merge into a very massive star and finally collapse to a BH with a few thousand solar masses \citep[e.g.,][]{portegies2004formation, omukai2008can, devecchi2009formation}. The third one is the direct collapse scenario, which starts from a supermassive star formed within the globally unstable galactic gas disks and leaves heavy BH seeds with up to $\sim$ one million solar masses \citep[e.g.,][]{begelman2006formation, agarwal2012ubiquitous}. Once formed, those seeds grow in mass via gas accretion and mergers through the hierarchical structure formation, and could produce observed high-$z$ quasars.

In order to disentangle the above different models, we first need to obtain the distributions of BH mass and Eddington ratio to quantify SMBH assembly in the early cosmic epoch. \citet{shen2019gemini} (S19) have reported  the properties of 50 quasars at $z \geq 5.7$ based on NIR spectra collected with the Gemini telescope. The S19 sample has the BH mass range of $\sim 10^{8-10} [M_{\odot}]$ with the average Eddington-ratio of 0.3, and no quasars have hyper massive SMBH ($M_{\rm BH}> 10^{10}\ M_{\odot}$) or super-Eddington ($\lambda _{\rm Edd}> 1$) accretion. Their sample has similar UV spectral properties to a low-$z$ luminosity-matched sample, both in the emission line profile and continuum.
\citet{yang2021probing}(Y21) have further studied the $M_{\rm BH}$ distribution of 37 quasars at $6.30 \leq z \leq 7.64$. The spectra were obtained with the Keck, Gemini, VLT, and Magellan telescopes. Most of their sample host massive SMBHs, which span the mass range of $M_{\rm BH} \sim$ (0.3--3.6)$\times 10^9 M_\odot$, resulting in the predicted seeds masses larger than $10^{3-4} M_{\odot}$ on the assumption of continuous Eddington accretion since $z= 30$. The Fe~{\sc ii}/Mg~{\sc ii} flux ratios are comparable to low-$z$ values and thus suggest the metal-rich environment.\\
However, the previous high-$z$ quasar studies are limited to the most luminous quasars with bolometric luminosity ($L_{\rm bol}$) $> 10^{46} \mathrm{erg\ s ^{-1}}$. 
We are currently carrying out a quasar survey project based on the Hyper Suprime-Cam (HSC) Subaru Strategic Program (SSP), a deep and wide-field survey with five broadband filters ($g$, $r$, $i$, $z$, $y$) plus narrow-band filters \citep{2018PASJ...70S...1M, 2018PASJ...70S...4A}. The HSC-SSP survey has three layers (Wide, Deep, and UltraDeep) with different combinations of area and depth. The Wide layer covers $1200 \mathrm{\ deg^2}$
while the UltraDeep layer has a  depth of $r_{\rm AB} \sim 28\ \mathrm{mag}$. Based on this survey,  our team, ``Subaru High-$z$ Exploration of Low-Luminosity Quasars (SHELLQs)'' project has constructed a sample of 162 low-luminosity quasars at $z\sim$ 6--7 \citep{matsuoka2016subaru, matsuoka2018subaru, matsuoka2019discovery, matsuoka2019subaru,matsuoka2022subaru} with the most distant object at $z = 7.07$ \citep{matsuoka2019discovery}. The luminosity range of the SHELLQs sample reaches about two orders of magnitude down the previous high-$z$ quasars sample, i.e., $M_{1450}\ < \mathrm{-21\ mag}$. This low luminosity sample is expected to include low-mass SMBHs at such high redshift. On the other hand, due to their faintness, most of the SHELLQs quasars are not detected in public near-IR surveys.

 \citet{2019ApJ...880...77O} present the measurements of Mg~{\sc ii} based BH mass of six relatively luminous ($M_{1450} \leq -24$ mag) SHELLQs quasars (three quasars also have measurements of C~{\sc iv} based BH masses), based on NIR follow-up spectroscopy. They found a wide black hole mass range of $M_{\rm BH}=10^{7.6-9.3} M_{\odot}$ and the Eddington ratio ($\lambda_{\rm Edd} = L_{\rm bol}/L_{\rm Edd}$, where $L_{\rm Edd}$ is Eddington luminosity) range of $\lambda_{\rm Edd} =$ 0.1--1.0,  but most black holes have masses of $\sim 10^9M_{\odot}$ and accrete with sub-Eddington ratios. It is difficult to reproduce the observed BH masses by keeping the observed sub-Eddington growth, so it implies much more active growth in the past. Thus, we may be witnessing their activity transition phase, changing from the active to quiescent mode. 
\citet{2018PASJ...70...36I,2019PASJ...71..111I,2021ApJ...908..235I,2021ApJ...914...36I} carried out ALMA observations of nine SHELLQs quasars and detected [C~{\sc ii}] ($\mathrm{158\ \mu m}$) emission line and infrared continuum. They show that the host galaxies of the SHELLQs sample are close to the main sequence of co-eval star-forming galaxies when they used ALMA dynamical mass as a surrogate for stellar mass, in contrast to more luminous high-$z$ quasars. The SMBH - host dynamical mass relation in the SHELLQs sample is close to the local SMBH-stellar mass relation, which may suggest that the co-evolution is already in place at $z\sim 6$. This is in clear contrast to luminous $z \sim 6$ quasars, whose BH masses are above the local relation. \\
This is the 17th paper from the SHELLQs project and is the first to present BH masses for a significant fraction of our sample. In this work, we tried to estimate their BH masses by comparing their spectra with a massive sample of low-$z$ quasar spectra. It also provides a new way to estimate the BH mass of high-$z$ quasars with reasonable accuracy, making it possible to discuss the evolutionary stage of the high-$z$ low luminosity quasars.\\
This paper is organized as follows. We describe the data and the method to get ``low-$z$ spectral counterparts'' in Section \ref{dataanalysis}. The main results are presented in Section \ref{sec:result}. The growth history of our sample and some other spectral properties are discussed in Section \ref{discussion}.
We adopt the cosmological parameters $H_{0} = 70\mathrm{\ km\ s ^{-1}\ Mpc^{-1},\ \Omega_{M} = 0.3,\ and\ \Omega_{\Lambda}= 0.7.}$ All magnitudes in the optical and NIR bands are presented in the AB system \citep{oke1983secondary}.

\section{Data \& Analysis}
\label{dataanalysis}

We exploit a huge sample of low-$z$ quasars and extract a ``counterpart'' to each of the SHELLQs quasars, with the following two assumptions. First, we assume that there are correlations between the spectral shape at $\lambda_{\rm rest}\sim$ 1200--1400 \AA, where $\lambda_{\rm rest}$ is rest-frame wavelength (observed by SHELLQs project), and that at the longer wavelengths which contain BH mass tracers, such as C~{\sc iv}, Mg~{\sc ii}, and H$\beta$ emission lines. The second assumption is that there exist one or more quasars in the low-$z$ universe that share the spectral shapes with the high-$z$ quasars.\\
We analyze 139 type-1 quasars at $5.66 \leq z \leq 7.07$ with rest-ultraviolet (rest-UV) absolute magnitudes $M_{1450} \sim -26 \mathrm{\ to} -21 \mathrm{\ mag}$ (median of $-$24 mag), constructed by the SHELLQs project. Their spectra have been obtained by two instruments: Faint Object Camera and Spectrograph (FOCAS; \citealp{2002PASJ...54..819K}) installed on the Subaru Telescope and the Optical System for Imaging and Low-Intermediate-Resolution Integrated Spectroscopy (OSIRIS; \citealp{2000SPIE.4008..623C}) installed on the Gran Telescope CANARIAS (GTC). The two spectrographs provide spectral coverage of $\lambda_{\rm obs} = 0.75 -1.05 \mathrm{\ \mu m}\mathrm{\ and\ } 0.74 - 1.0 \mathrm{\ \mu m}$ (where $\lambda_{\rm obs}$ is observed-frame wavelength), respectively. The spectral resolution ($R$) $\sim 1500 \mathrm{\ and\ }1200$, respectively. Our discovery spectra cover only a small portion of the rest-UV wavelength at $\lambda_{\rm rest} \sim$ 1200  to 1400 \AA, and thus do not include emission lines to derive black hole masses, e.g., C~{\sc iv} and Mg~{\sc ii} lines. 

We have used SDSS Data Release fourteen (DR14) quasar catalog \citep{2018A&A...613A..51P} as a low-$z$ quasar sample to extract the counterpart spectra. The catalog contains $\sim $520,000 quasars discovered over a quarter of the all-sky ($\sim 10000 \mathrm{\ deg^2}$) observed through SDSS-I/II/III/IV with various selections. For example, the targets of SDSS-IV were selected by multiple selection algorithms using X-ray, optical, infrared, and radio data. X-ray sources were observed in SDSS-IV/SPIDERS (SPectroscopic IDentification of ERosita Sources). On the other hand, SDSS-IV/eBOSS selected targets based on three imaging data: SDSS, \textit{Wide-field Infrared Survey Explorer} (\textit{WISE}: \citealt{2010AJ....140.1868W}), and the Palomar Transient Factory (PTF; \citealt{2009PASP..121.1334R}; \citealt{2009PASP..121.1395L}).
Quasar candidates for SDSS-IV/eBOSS CORE sample were selected with morphological requirements and optical and\textit{WISE} color cut based on XDQSOz method \citep{2012ApJ...749...41B}. At higher redshift ($z > 2.1$), candidates were selected using their PTF photometric variability. In addition, known quasars that have low-quality spectra in SDSS-III/BOSS were re-observed in SDSS-IV.
SDSS-IV/eBOSS also selected candidates from sources within $1 \arcsec$ of a radio detection in the FIRST point source catalog \citep{1995ApJ...450..559B}, and also selected ``time domain spectroscopic survey'' targets in the $g,r$ and $i$ bands using the SDSS-DR9 imaging data \citep{2012ApJS..203...21A} and the multi-epoch Pan-STARRS (PS1) photometry \citep{2002SPIE.4836..154K,2010SPIE.7733E..0EK}. The SDSS-IV targets were observed by the eBOSS spectrographs whose resolution varies from $\sim$ 1300 at 0.36 $\rm \mu m$ to 2500 at 1.0 $\rm \mu m$ \citep{2013AJ....146...32S}, in a series of at least three 15-min exposures. Since the SHELLQs quasars are likely a mixture of radio-loud and quiet quasars, we didn't remove radio-loud quasars from the SDSS sample when selecting counterparts. We also didn't remove BAL quasars from SDSS, since the SHELLQs sample clearly includes BAL quasars (see below).

The redshift and rest-UV magnitude ranges of the SDSS sample are 0 $<$ z $<$ 5 and $M_{i}[z = 2] < -20.5$ mag, where $M_i[z=2]$ represents $i$-band absolute magnitude at $z=2$. We selected the sample with the flag ``ZWARNING==0'' and removed quasars that have ``$\mathrm{LOG\_LYA= -999.0}$''.
100,888 quasars at $2.5 \leq z \leq 5.0$ remain with the above criteria as a parent sample of counterparts in this work; the lower redshift cut was incorporated to include Ly$\alpha$, which is the only characteristic emission line among the obtained spectra of the SHELLQs sample. Most of these low-$z$ quasars are at $z\sim$ 2.5--3.0 and thus, we used their C~{\sc iv} lines to estimate the BH mass of SHELLQs quasars.

\subsection{Extraction of ``low-z counterparts''}\label{subsec:counterparts}
We performed spectral fitting in $\lambda_{\rm rest} \sim$ 1200--1400 \AA \ between a SHELLQs quasar and the low-$z$ sample via $\chi^2-$fitting, 

\begin{eqnarray}
   \label{chi2fitting}
   \chi^2 = \sum{i} \frac{(f_{i}^{SHELLQs}-Af_{i}^{SDSS})^2}{\sigma_i^2}  
\end{eqnarray}

where $i$ represents the flux data points, $f^{\rm SHELLQs}_i$ and $f_{i}^{\rm SDSS}$ are the SHELLQs flux density and the SDSS flux density, respectively. $A$ is a free parameter. The flux errors of the low-$z$ quasars are much smaller than those of the SHELLQs quasars, and thus we considered only the errors of the latter spectra as $\sigma_{i}$ in Equation(\ref{chi2fitting}). Ideally, we should select a counterpart with simillar luminosity to each of the SHELLQs quasars, but this is practically impossible since the SHELLQs quasars are significantly less luminous than the SDSS quasars at $z$ > 2.5. We also tried to vary redshift as a free parameter, but it does not improve the results and the additional flexibility sometimes leads to catastrophic fits. Redshifts are therefore treated as the fixed values.

\begin{figure*}[t]
\begin{center}
\includegraphics[height = 16cm, width = 18cm]{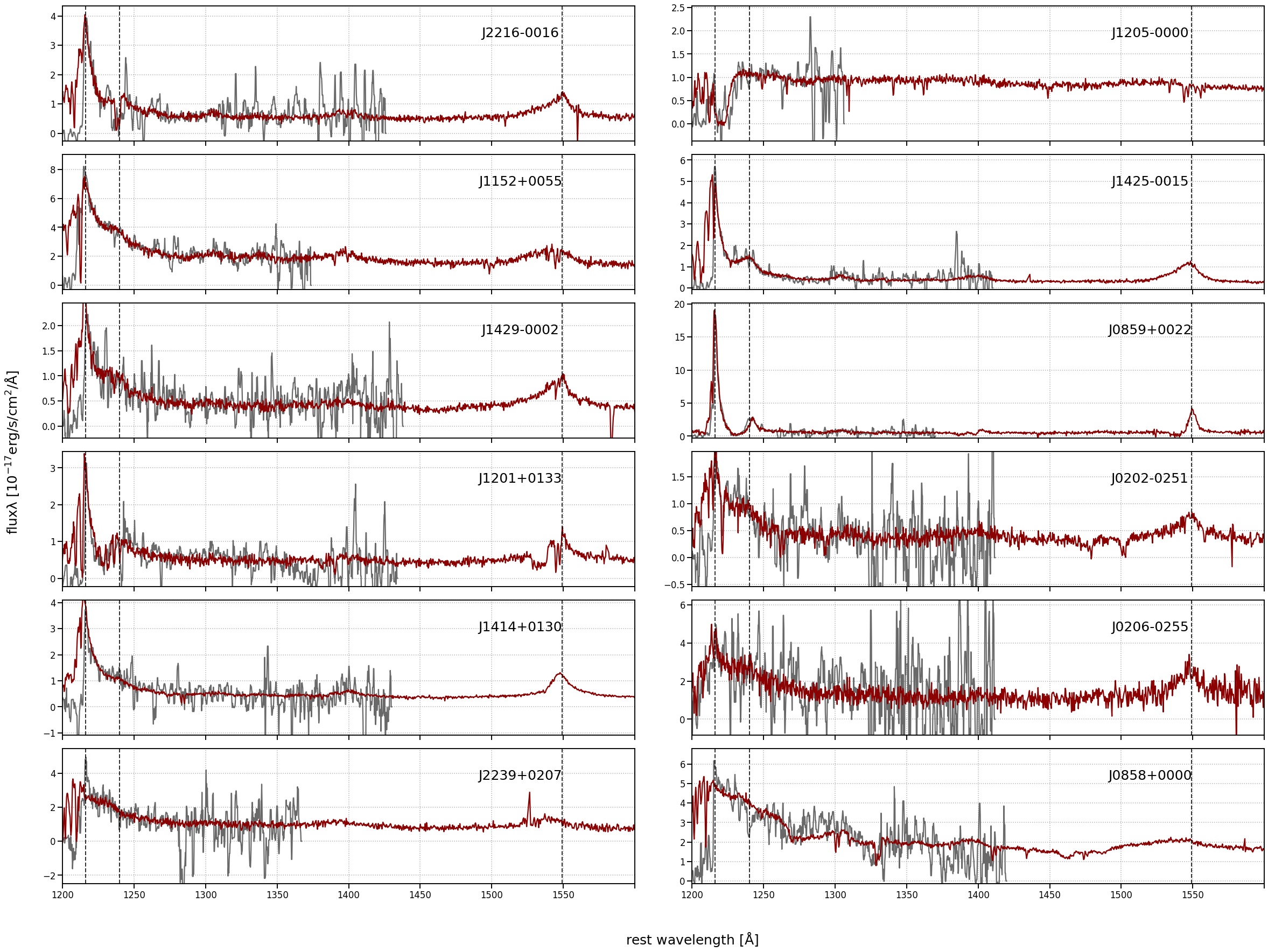}
\caption{Example of the spectral fitting between SHELLQs quasars (gray solid line) and their low-$z$ counterparts from SDSS (red solid line). The dashed lines mark the expected positions of  Ly$\alpha$, N~{\sc v}, and C~{\sc iv} emission lines. \label{fig:ex_cp}}
\end{center}
\end{figure*}

We extract the SDSS quasar with the most similar shape to each SHELLQs quasar. We performed $\chi^2-$fitting at the redward of $\mathrm{Ly\alpha}$ emission line since the bluer part is severely affected by the IGM absorption for the high-$z$ quasars. Figure \ref{fig:ex_cp} shows examples of the fitting results.\footnote{Fitting results of all the 139 quasars are presented at \url{https://cosmos.phys.sci.ehime-u.ac.jp/~takahashi/figures/spec.pdf}} The spectral shapes of the low-$z$ counterparts are indeed very similar to the features of the SHELLQs quasar spectra, not only in emission but also in absorption features; this is evident particularly in, e.g., J2216$-$0016, J1205$-$0000, J1201$+$0133. As we will see below, the counterpart spectra have proven to reproduce the actual spectra even at the longer wavelengths, for a limited number of objects whose NIR spectra are already available. This means that the Ly$\alpha$ and other spectral features in 1200 -- 1400 \AA\ are indeed correlated with the spectral shape at $>$ 1400 \AA \ which includes the BH mass tracers. \\

\subsection{How to measure BH masses}\label{estimates_mbh}

We estimate SMBH mass with a single epoch method based on the virial relation, using the calibration relation
\begin{equation}
 \begin{split}
   \label{BHmass}
   \log\frac{M_{BH}}{M_\odot} = A + B\log\frac{\lambda L_{\lambda}}{\mathrm{10^{44}ergs^{-1}}}\\
            + 2\log\frac{\mathrm{FWHM}}{\mathrm{kms^{-1}}} 
\end{split}
\end{equation}

, where FWHM is the full-width at half maximum of the C~{\sc iv} line, and $\lambda L_{\lambda}$ is the monochromatic luminosity at 1350 \AA, radiated from the accretion disc. This equation is based on the tight relation between the radius of the line-emitting region and the continuum luminosity seen in the local universe \citep[e.g.,][]{2005ApJ...629...61K}. We adopt (A, B) = (0.66, 0.53) following \citet[VP06]{2006ApJ...641..689V}.

\begin{figure*}[h]
\begin{center}
\includegraphics[height = 15cm, width = 18cm]{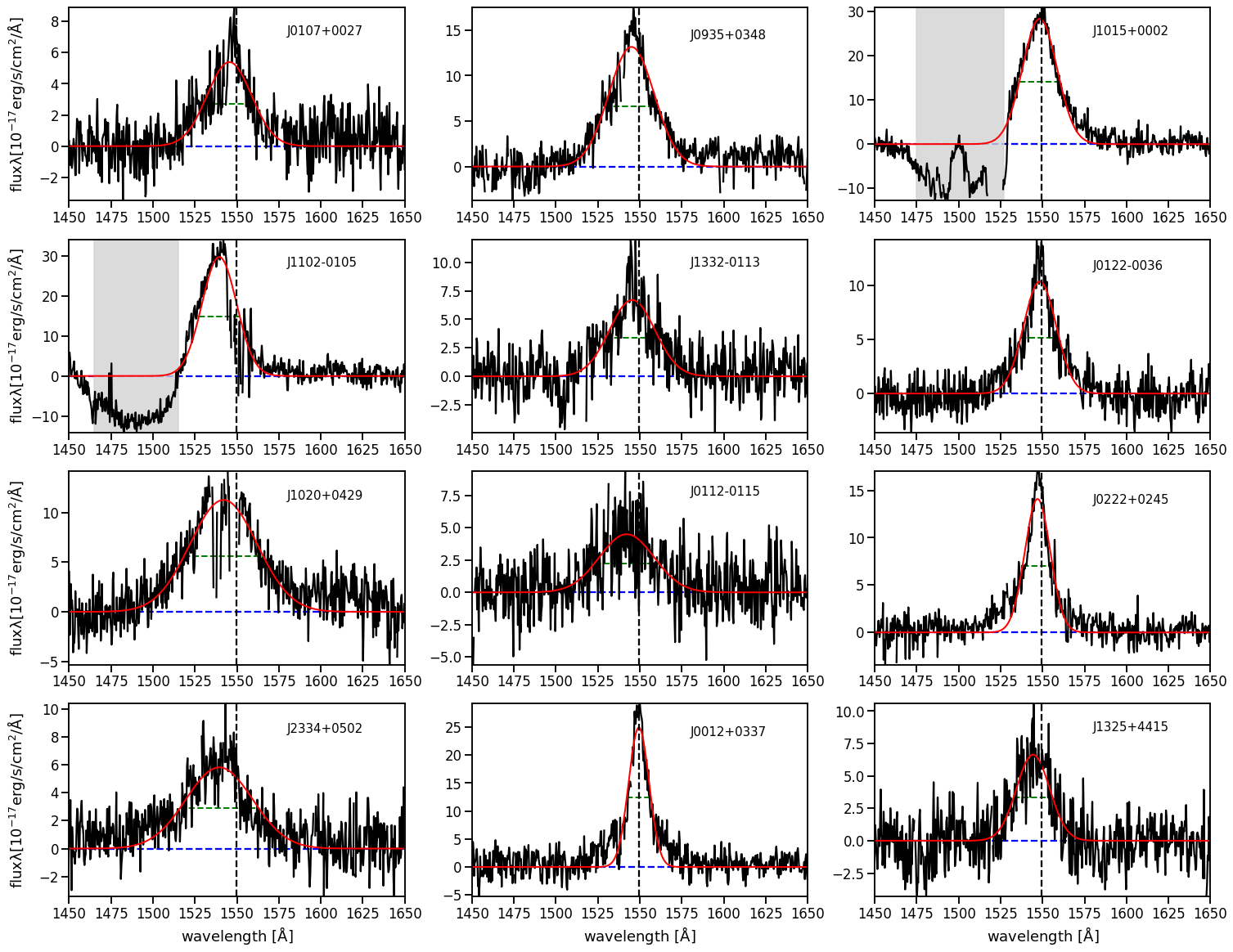}
\caption{The best-fit models of continuum-subtracted spectra around the C~{\sc iv} lines. Black solid lines show the observed flux data, and gray shaded areas represent the masked region. The best-fit Gaussian profiles are in red, and FWHMs are in green.
Black dashed lines represent the theoretical peak wavelength of 1549 \AA. \label{fig:ex_fwhm}}
\end{center}
\end{figure*}

\begin{figure}[htbp]
\begin{center}
\includegraphics[width=7cm,height=7cm]{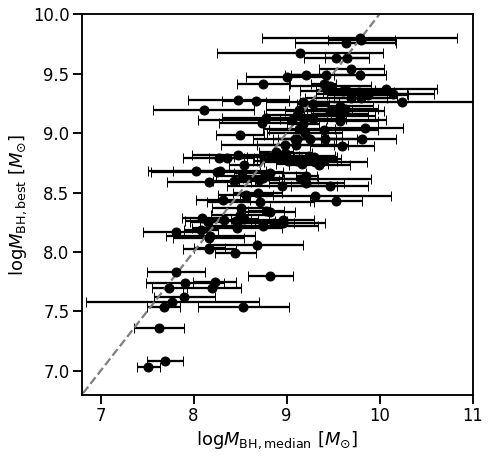}
\caption{Comparison of BH masses measured from one counterpart that has the smallest reduced-$\chi^2$ (in the vertical axis) with those measured from ten counterparts (in the horizontal axis). The horizontal error bars represent 1 sigma scatter of the ten counterparts.  \label{fig:best_median}}
\end{center}
\end{figure}

We used the observed luminosity of each SHELLQs quasar to calculate $\lambda L_{\lambda}$, calculated from $M_{1450}$ assuming a power-law continuum slope of $-1.5$. The C~{\sc iv} line widths were measured from the counterpart quasars, as follows. We first subtracted the continuum emission estimated at both sides of the emission line. Next, we masked broad absorption features identified visually and removed outlying flux data points with sigma-clipping. 
Finally, we fitted C~{\sc iv} emission lines by varying the scaling factor, central wavelength, and width of single or two Gaussian profiles. The fitting wavelength range is 1450 --1650 \AA. The line FWHMs were measured with those best-fit models.
The uncertainties were measured with 1000 mock spectra for each counterpart generated by adding random noise to each spectral pixel based on the observed noise vectors.
Figure \ref{fig:ex_fwhm} shows examples of our fitting results. 
We visually inspected all the fitted spectra and excluded eight quasars with poor fits, likely due to large flux errors and/or strong absorption features, from the BH mass estimates presented below.
Once we estimate BH mass, the Eddington ratios are straightforward to measure. We derive the bolometric luminosity ($L_{\rm bol}$) with  $L_{\rm bol} = 3.81 \times L_{\rm 1350}$ \citep{2006ApJS..166..470R} where $L_{\rm 1350}$ is a monochromatic luminosity at 1350 \AA. In reality the bolometric correction varies from one object to another, so we must keep in mind that this simple scaling from $L_{\rm 1350}$ give rise to additional uncertainty in the bolometric luminosity and Eddington ratio. \\
We also tested obtaining 10 counterparts (with the smallest reduced-$\chi^2$ values) and estimating BH mass as the median of the 10 measurements for each SHELLQs quasar. The comparison between this 10-counterparts case with our default case (one counterpart with the least reduced-$\chi^2$) is shown in Figure \ref{fig:best_median}. Overall, the two sets of measurements are consistent with each other. For simplicity and limited computing resources, we use the one-counterpart case in the following analysis and discussions. 

\subsection{Uncertainty of C~{\sc iv} based BH measurement}
C~{\sc iv} is one of the most prominent broad BLR emission lines commonly used as BH mass estimators for $z \geq 2$ quasars. However, it frequently has asymmetric profiles, broad absorptions at the blue sides of their peaks, and /or blueshift in the line centroid, which are in clear contrast to lower ionization lines such as H$\beta$ and Mg~{\sc ii} \citep[e.g.,][]{2000ARA&A..38..521S, 2011AJ....141..167R, brotherton2015bias, 1982ApJ...263...79G, 1992ApJS...79....1T}. Such features suggest that C~{\sc iv} is more severely affected than other lines by nonvirial gas motion, which would also depend on the viewing angle. The fiducial recipes of BH mass estimates summarized in \citet{shen2011catalog} are known to have substantial systematic uncertainty of $\sim 0.4$ dex, estimated from the differences among different broad-line estimators, such as C~{\sc iv} and Mg~{\sc ii}. This systematic uncertainty is often larger than the measurement uncertainty of spectral fits. In contrast, Mg~{\sc ii} line widths are well correlated with those of H $\beta$, and these two low-ionization line estimators usually give consistent virial masses \citep[ e.g.,][]{2004MNRAS.352.1390M, 2007ApJ...662..131S}.
\citet{shen2008biases} compiled BH masses of $\sim 60,000$ quasars in the redshift range $0.1 \leq z \leq 4.5$ based on the H$\beta$, Mg~{\sc ii}, and C~{\sc iv} emission lines. 
Within their sample, the Mg~{\sc ii} and H$\beta$ based BH masses are on average consistent with the C~{\sc iv} estimates.
However, they found a systematic offset between the C~{\sc iv} and Mg~{\sc ii} BH masses.
The offset correlates with the blueshift of C~{\sc iv} relative to Mg~{\sc ii}, which suggests that the C~{\sc iv} line is more severely affected by a disk wind. 
Particularly in high luminosity quasars, C~{\sc iv} has been presumed to be a non-ideal line to measure BH masses, while it is the only line that can be used to estimate BH masses at high redshifts \citep[e.g.,][]{2005MNRAS.356.1029B, 2013BASI...41...61S,2017MNRAS.465.2120C}.
On the other hand, \citet{2019ApJ...880...77O} compared the Mg~{\sc ii}- and C~{\sc iv}-based BH masses of three high-$z$ low-luminosity quasars and found that the two estimates show good agreement with each other. This may indicate that the C~{\sc iv} emitting region of lower-luminosity quasars may be less affected by gas outflows, which may be weaker. In what follows, we assume that the BH mass distributions presented below are not subject to strong systematic uncertainties while keeping all the above facts in mind. We also tested the BH mass calibration that takes into account and corrects for the effect of C~{\sc iv} blueshift component, presented by \citet{2017MNRAS.465.2120C}. This calibration was empirically established from 230 high-luminosity ($ L_{\rm bol} \sim 10^{45.5-48.5} {\rm erg\ s^{-1}}$) quasars at $1.5 < z < 4.0$, which covers both the hydrogen Balmer emission lines and the C~{\sc iv} emission lines. We found little change in the estimated BH masses, and thus confirmed that the conclusions of this paper remain unchanged with this alternative calibration.

\subsection{Consistency of our BH mass estimates with direct measurements}
\begin{figure}[htbp]
\begin{center}
\includegraphics[width=7cm,height=21cm]{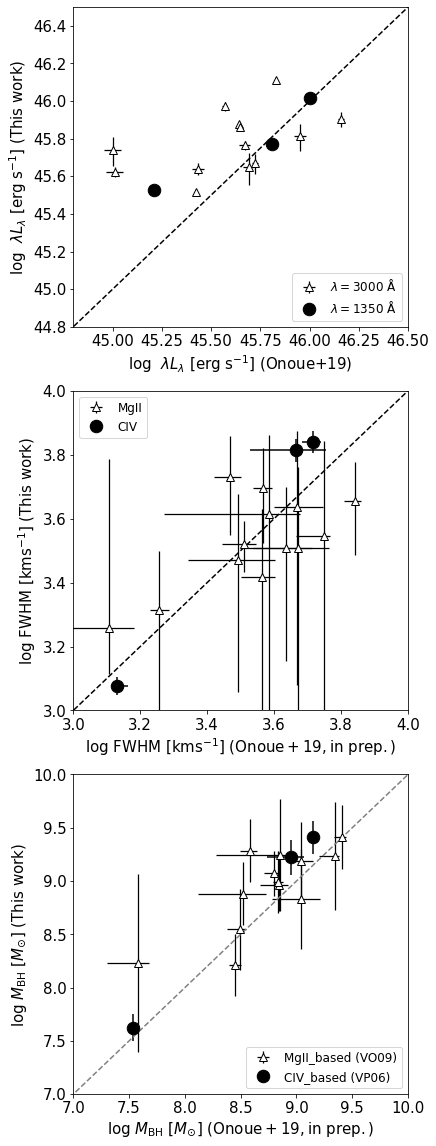}
\caption{Comparisons between our estimates (vertical axis) and actual measurements (horizontal axis; Onoue et al. 2019, M. Onoue et al., in prep.) of the continuum luminosity (top), line FWHM (middle), and BH mass (bottom). The dots and triangles represnt the C~{\sc iv}-based and Mg~{\sc ii}-based estimates, respectively.
\label{fig:comp_actualdata}}
\end{center}
\end{figure}

\begin{figure}[htbp]
\begin{center}
\hspace*{-0.5cm}
\includegraphics[width=9cm,height=6cm]{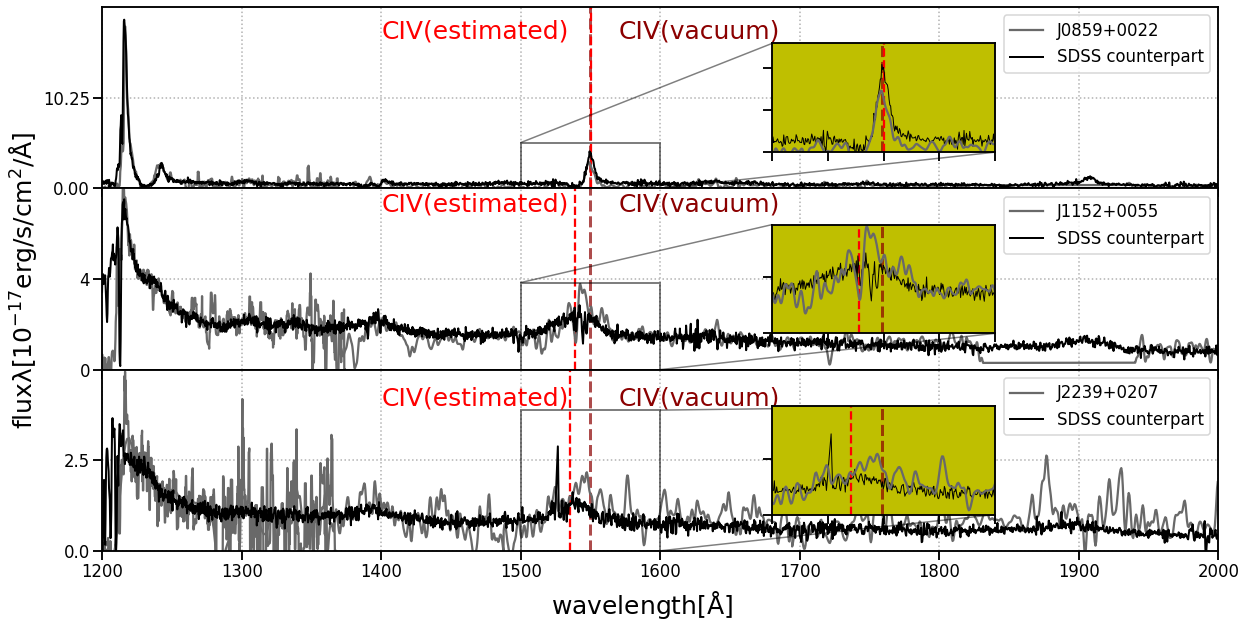}
\caption{Spectral comparison of three quasars with C~{\sc iv}-based BH mass estimates (corresponding to the dots in Figure \ref{fig:comp_actualdata}). Gray solid lines show the NIR spectra obtained by O19.  Black solid lines show the low-$z$ counterpart spectra. Red and brown lines mark the peak wavelengths estimated by our Gaussian fitting and theoretical wavelength of the C~{\sc iv} lines, respectively. Yellow windows provide the extended plots around C~{\sc iv} (at 1500 --1600 \AA). The two sets of spectra (SHELLQs vs. counterparts) are in reasonable agreement with each other, even outside of the spectral coverage ($>$ 1400 \AA) used for the fitting. \label{fig:spec_civ_onoue}}
\end{center}
\end{figure}

Here, we compare our BH mass estimates with those based on the direct spectral measurements provided by \citet[O19]{2019ApJ...880...77O}.
O19 collected NIR spectra of six SHELLQs quasars at $6.1 < z < 6.7$ and with $M_{1450} \leq -24$ mag, which is a practical limit to obtain reasonable emission line measurements with, say, less than 10 hours with 8-m class telescope on the ground. Specifically, O19 observed the six quasars with Very Large Telescope/X-Shooter and Gemini-N/GNIRS, with the spectral resolution of $R\sim$ 5000 -- 7000 and $R\sim$ 500 --800, respectively. The resultant Mg~{\sc ii} ($\lambda 2798$) based BH masses are in the range of $M_{\rm BH} = 10^{7.6 - 9.3} M_{\odot}$, with the Eddington ratio of $L_{\rm bol}/L_{\rm Edd} = 0.16-1.1$, while most of the quasars have $M_{\rm BH} \sim 10^9 M_{\odot}$ with sub-Eddington accretion. Three quasars in the O19 sample also have $M_{\rm BH}$ estimates based on C~{\sc iv} emission lines. In addition, nine more quasars from the SHELLQs sample has the observed NIR spectra to measure BH masses (M. Onoue et al., in prep.). \\
Figure \ref{fig:comp_actualdata} shows comparisons of the continuum luminosities, line widths (FWHM), and BH masses between the direct spectral measurements (horizontal axis) and the present work based on the low-$z$ counterparts (vertical axis).
In addition to the C~{\sc iv}-based measurements, we include Mg~{\sc ii}-based measurements in this comparison; the counterparts for the latter have been newly selected from SDSS in a limited redshift range ($2.1 < z < 2.5$) where both Ly$\alpha$ and Mg~{\sc ii} are covered by the SDSS spectroscopy, and the BH masses have been estimated from their Mg~{\sc ii} in exactly the same way as we did based on C~{\sc iv}.
 These comparisons suggest that our method of estimating $M_{\rm BH}$ via low-$z$ counterparts works well with reasonable accuracy. 
While there are a modest amount of scatters in the plotted quantities, the BH masses from the two methods are consistent with each other within $\sim 0.3$ dex, which is comparable to the typical systematic uncertainty expected in the single-epoch mass estimates.\\
Figure \ref{fig:spec_civ_onoue} represent the spectral comparisons of the three quasars and their counterparts with C~{\sc iv}-based $M_{\rm BH}$ (i.e., those corresponding to the dots in Figure \ref{fig:comp_actualdata}). While there is a good overall agreement between the two sets of spectra, we observed offsets in C~{\sc iv} peak wavelengths of J1152$+$0055 and J2239$+$0207 from the expected positions. This may be caused by uncertainty in redshift of the SHELLQs quasars, which have been determined from the Ly$\alpha$ line affected severely by the IGM absorption.

Since [C~{\sc ii}] redshifts are available for the above three quasars from SHELLQs ALMA observations \citep{2018PASJ...70...36I,2019PASJ...71..111I}, we tested to use those more accurate redshift to convert observed spectra into the rest-frame and investigate how the redshift uncertainly affects the accuracy of this counterpart method. 
The counterparts and thus BH mass estimates of two out of the three quasars have changed from the case with Ly$\alpha$ redshift, as shown in Figure \ref{fig:mbh_calbz}
 (see Figure \ref{fig:spec_civ_calbz} in Appendix for the updated spectral comparison similar to Figure  \ref{fig:spec_civ_onoue}). We observe almost perfect agreement between the two sets of measurements in this case, demonstrating the potential power of the present method. However, counterparts with Ly$a$ redshifts (the only redshift estimates available for the majority of the SHELLQs quasars) still provide reasonable BH mass estimates with $\sim 0.2$ dex difference from the [C~{\sc ii}] redshift cases, and such small difference does not affect the conclusion of this paper. Having said that, we stress that future follow-up observations of those quasars to measure more accurate redshifts, with [C~{\sc ii}], Mg~{\sc ii} and/or other emission lines, will significantly enhance the usefulness of this unique high-$z$ quasar sample for various topics.\\

\begin{figure}
    \centering
    \includegraphics[width=7cm,height=7cm]{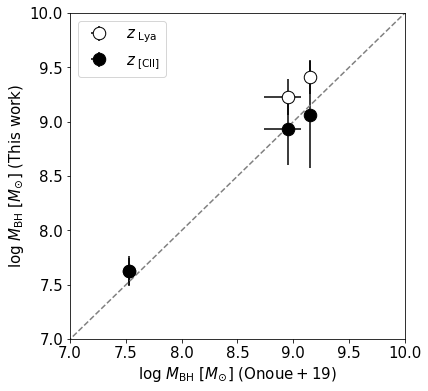}
    \caption{Same as Figure \ref{fig:comp_actualdata} (bottom), but for C~{\sc iv}-based BH mass estimates with Ly$\alpha$ redshifts (open circles) and with [C~{\sc ii}] redshifts (dots).}
    \label{fig:mbh_calbz}
\end{figure}

\begin{figure}[htbp]
\begin{center}
\includegraphics[width=7cm,height=7cm]{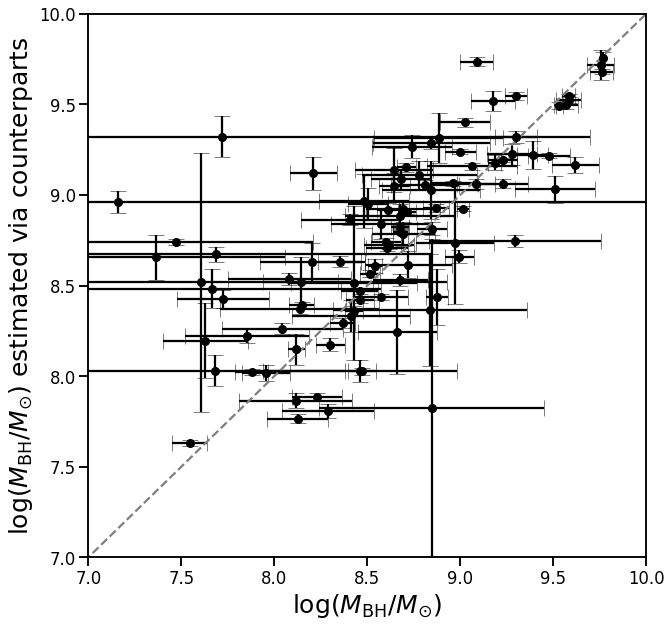}
\caption{Comparison of BH masses between direct measurements and counterpart estimates for a randomly selected sample of 100 SDSS quasars at $z \sim$ 2--3.  \label{fig:randomcheck}}
\end{center}
\end{figure}

As an additional test of the present method, we randomly selected 100 quasars from $z\sim$ 2--3 SDSS sample and cutout their spectra in the range of $\lambda_{\rm rest}=$1200 -- 1400 \AA, and then obtained their counterparts from the remaining SDSS quasars in exactly the same way as we did for the SHELLQs quasars. Figure \ref{fig:randomcheck} compares BH masses based on direct measurements from C~{\sc iv} lines contained in the original 100 spectra and those based on the counterpart method. We found a broad agreement between the two measurements, with the systematic offset and standard deviation of 0.09 dex and 0.23 dex, respectively; these results show a good correlation even on the high-mass side.

\begin{figure}[htbp]
\begin{center}
\includegraphics[width=7cm,height=7cm]{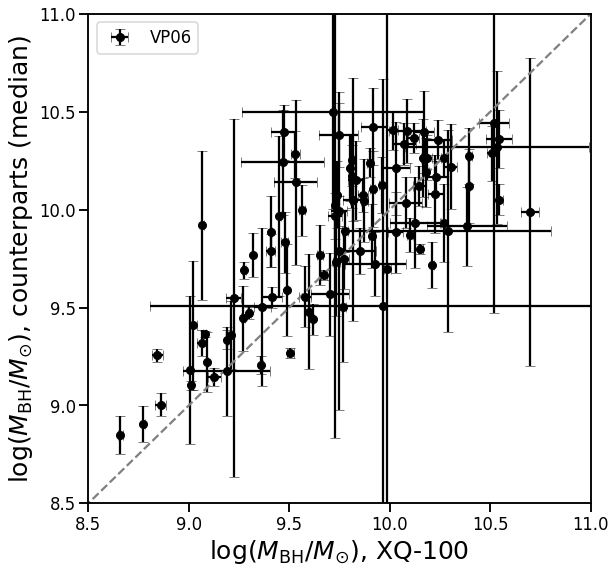}
\caption{Comparison of BH masses between direct measurements and counterpart estimates for XQ-100 quasars $z\backsimeq 3.5-4.5$.  \label{fig:xq-100}}
\end{center}
\end{figure}

We further tested our method with luminous quasars taken from the XQ-100 sample.
This sample was constructed based on the European Southern Observatory Large Programme (program code 189.A-0424) ``Quasars and their absorption lines: a legacy survey of the high-redshift Universe with VLT/X-shooter''. The XQ-100 survey produced high-resolution ($R\sim$4000--7000) spectra of 100 quasars at redshift $\simeq$ 3.5--4.5 covering Ly$\alpha$ and C~{\sc iv} emission lines \citep{2016A&A...594A..91L}. The median absolute magnitude is $M_{\rm 1450}=$ -29.6 $\pm$ 0.017 and the median bolometric luminosity is $L_{\rm bol} = (8.0\pm 0.13) \times 10^{47}\rm erg s^{-1}$. We measured their BH mass using C~{\sc iv} lines with the VP06 calibration, and calculated their uncertainties based on the MCMC method with 100 mock spectra. At the same time, we estimated thier BH mass with the counterpart method using the SDSS spectra. The comparison between the two measurements is shown in Figure \ref{fig:xq-100}. We found a reasonable agreement again, with the systematic offset and standard deviation of 0.13 dex and 0.22 dex, respectively. Overall, we found that our new method provides a reasonable accuracy of BH mass estimates for high-$z$ quasars, whose near-IR spectra are not easy to obtain, especially for those with low luminosities.

\section{BH mass Distribution of high-z low-luminosity quasars} \label{sec:result}

\begin{figure*}[ht]
\begin{center}
\includegraphics[height = 10cm, width = 10cm]{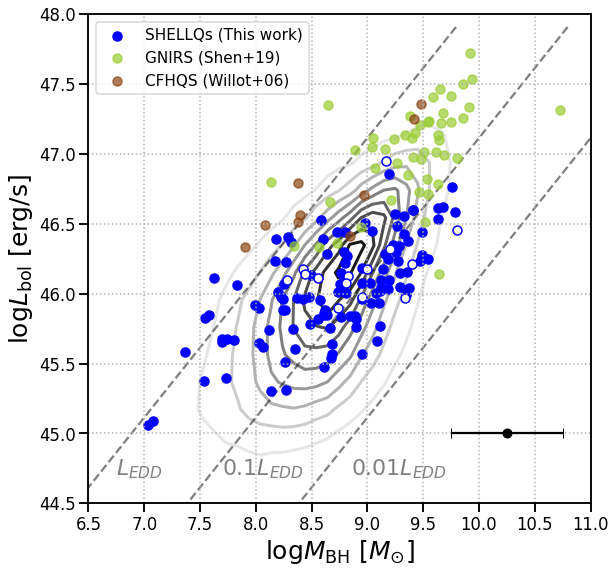}
\caption{The SMBH mass-luminosity plane of  high-$z$ ($z > 5.6$) quasars. Blue dots represent our sample, whose BH masses were measured with the counterpart method. Blue open circles represent those with possible BAL features in the counterparts, with the SDSS measured BALnicity index BI $>0.0$ \citep{2018A&A...613A..51P}. The green and brown dots represent the samples from \citet{shen2019gemini} and \citet{2010AJ....140..546W}, respectively. The contours show the distribution of SDSS DR14 quasars at lower redshifts. The diagonal lines show the constant Eddington ratios of $L_{\rm bol}/L_{\rm Edd} = 1, 0.1, 0.01$ from top left to bottom right. The typical systematic uncertainty of the $M_{\rm BH}$ measurements (0.5 dex) is shown with the error bar at the lower right.} \label{fig:MBH}
\end{center}
\end{figure*}

\begin{figure}[htbp]
\begin{center}
    \includegraphics[height = 5cm,width = 7cm]{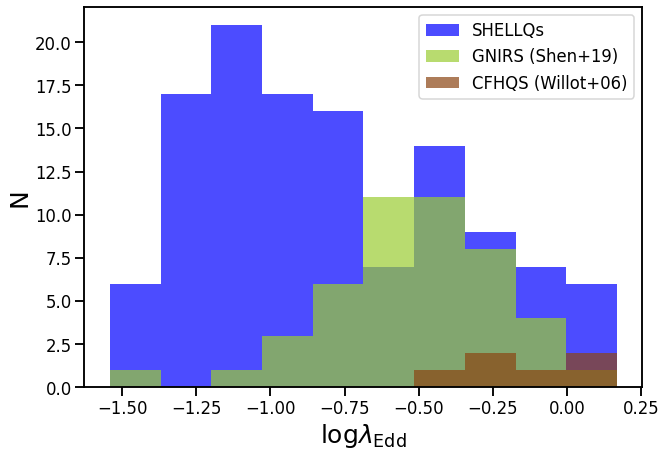}
\caption{The Eddington ratio distribution of SHELLQs (blue), CFHQS (brown), and GNIRS (yellow-green) samples.}
\label{fig:dist_Eddingtonratio}
\end{center}
\end{figure}

We show the relationship between the BH masses and bolometric luminosity of the SHELLQs sample in Figure \ref{fig:MBH}. Single epoch estimates of BH masses for high-$z$ quasars, based on calibrations established at lower redshift, are known to have large systematic uncertainties, whose typical value of 0.5 dex is indicated at the lower right of the figure. Our high-$z$ and other samples positively correlate in this parameter space. The estimated BH masses and the bolometric luminosities span a wide range of $M_{\rm BH} \sim 10^{7-10}\ \rm M_{\odot}$ and $\sim 10^{45.0-47.0} \mathrm{\ erg/s}$, with the median values of $\sim 10^{8.6}\rm M_{\odot}$ and $\sim 10^{46.1} \mathrm{\ erg/s}$, respectively.
For comparison, we also plot the quasar sample at lower redshift, SDSS DR14 quasars with BH masses estimated using either C~{\sc iv}, Mg~{\sc ii} or $\mathrm{H \beta}$ depending on redshift in \citet{2020ApJS..249...17R}. 
The BH masses of the luminous S19 sample at  $z \sim 6$ are based on either C~{\sc iv} or Mg~{\sc ii} and populate the range between $10^{8-10} \mathrm{\ M_{\odot}}$ except for one quasar at $>10^{10.5} \rm M_{\odot}$ with poor spectral fit. Our BH mass and Eddington ratio are lower on average by $\sim1$ dex and $\sim 0.15$, respectively than those in the S19 sample.
On the other hand, \citet{2010AJ....140..546W} (W10) reported Mg~{\sc ii}-based BH masses of nine high-$z$ quasars with relatively low luminosity ($M_{1450} < -24.3$ mag). 
They have $M_{\rm BH}\sim 10^{8-10} M_{\odot}$ with sub-Eddington to Eddington accretion.
The median BH masses of W10 are not so different from our BH masses, but the median Eddington ratio is about seven times higher than our sample. Sub-Eddington accreted quasars dominate our sample, representing the unprecedented depth of our survey. Figure \ref{fig:dist_Eddingtonratio} presents a histogram of Eddington ratios in the three sample, clearly showing the dominance of sub-Eddington accretors in the SHELLQs quasars. This is consist with the prediction from a semi-analytic models of galaxy formation, reported by  \cite{2019MNRAS.487..409S}. They calculated the Eddington ratio distributions of SMBHs with varying sample selection at 0 < $z$ < 6, and found that the shallower luminosity cuts tend to miss AGNs with lower Eddington ratios, and that such a trend is more critical at higher redshit.

We found six quasars marginally exceeding the Eddington limit. J0859$+$0022 has been observed by O19, who reported the Eddington ratio of $1.1^{+0.5}_{-0.3}$. Figure \ref{fig:spec_superedd} in Appendix shows the spectrum of one of the SHELLQs quasars with the highest estimated Eddington ratios. Their spectra tend to have high equivalents widths of Ly$\alpha$ and C~{\sc iv}, the latter being observed in the counterparts.
Similarly, lower-mass ($M_{\rm BH} < 10^{8} M_{\odot}$) quasars in our sample tend to have high EW of the emission lines. Figure \ref{fig:spec_lowmass} shows the spectrum of our lowest mass quasar along with that of its counterpart. Its estimated BH mass is $10^{(7.0\pm 0.2)} M_\odot$ and the Eddington ratio is 0.8. The high EW of the emission lines in these quasars may indicate gas-rich environments around the nuclei. Given the high Eddington ratios and/or low BH masses, they may represent the early stage of the assembly, accompanied by abundant inflowing gas. Figure \ref{fig:MBH} suggests that there are few hyper massive SMBHs ( $>10^{10} M_{\odot}$) in the present and other samples.
This is partly due to the observing difficulties since broad emission lines become extremely broad and weak compared to the continuum in such objects. Alternatively, we could think of a possibility that the AGN timescale at the high-mass end is so short that we cannot find them in our surveys over limited portions of the sky. It is also possible that the BH mass limit is actually determined by the host galaxy masses. \\

\section{Discussion}
\label{discussion}

\subsection{SMBH growth history}
The growth of a SMBH is exponential if it keeps a constant Eddington ratio. The timescale for a seed black hole with mass $M_{\rm seed}$ to reach a given $M_{\rm BH}$ is 
\[t_{\rm grow} = \tau \mathrm{ln}\frac{M_{\rm BH}}{M_{\rm seed}}\]
where $\tau$ is the e-folding timescale,
\[ \tau = 0.45 \left( \frac{\eta}{1-\eta}\right) \left( \frac{L_{\rm bol}}{L_{\rm Edd}} \right)^{-1} \mathrm{Gyr.}\]
The radiation efficiency $\eta$ is the factor describing how efficiently the accreting mass is converted to radiation. 

\citet{2018Natur.553..473B} reported the possible growth history of three luminous quasars at $z = 6.3 - 7.5$. They found $10^{3} < M_{\rm seed} < 10^{4} M_{\odot}$ at $z = 30$, assuming that the quasars had been accreting at the Eddington limit with a radiative efficiency of $10\%$. This suggests that SMBHs in the early universe have large initial masses and/or sustained Eddington limit (or episodically super-Eddington) accretion. The most distant quasar currently known, J0313-1806 at $z = 7.642$ \citep{wang2021luminous}, has the most massive black hole ($M_{\rm BH}\sim 10^{10}  M_{\odot}$) at $z > 7$, and thus poses a powerful constraint on the seed black hole mass. The estimated seed mass is $ M_{\rm seed} \sim 10^{4-5}  M_{\odot}$ with the assumptions of Eddington-limited accretion with a $10\%$ radiative efficiency and a duty cycle of unity. The above results are contrary to the scenario assuming the seeds of Population III star remnants.

\begin{figure}[htbp]
\begin{center}
    \includegraphics[height = 5cm,width = 8cm]{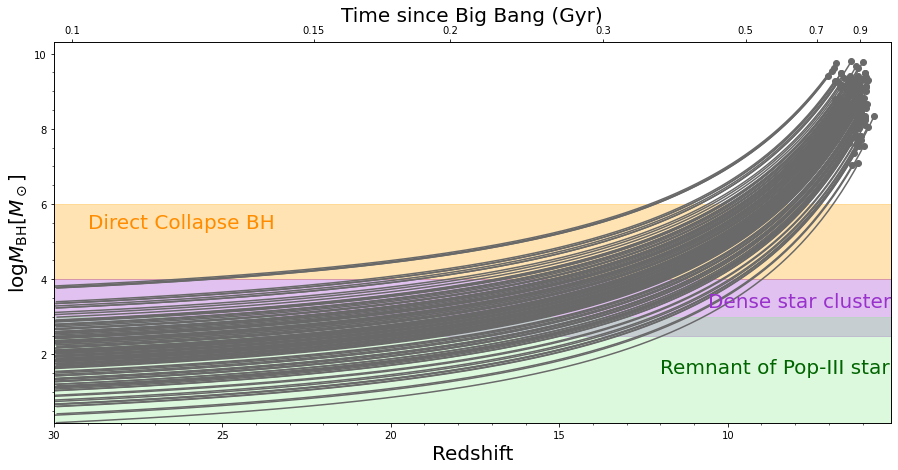}
    
    \caption{An estimated growth history of the SHELLQs quasars. The horizontal axis gives redshift (bottom) and time since the Big Bang (top). Solid lines show the cases where SMBHs have grown at the Eddington limit $\lambda_{Edd} = 1$, with the dots representing the estimated BH masses at the observed redshift. The shaded regions correspond to the predicted typical mass ranges for the three scenarios, i.e., Pop-III remnants ($M_{\rm BH} \leq 10^{3} M_{\odot}$; green), dense star clusters ($M_{\rm BH} \sim 10^{3-4}M_{\odot}$;  purple) and direct collapse BHs ($M_{\rm BH} \sim 10^{4-6}M_{\odot}$; orange).  \label{fig:growth__1}}
    \end{center}
\end{figure}

Here, we trace back in time the BH masses of the 131 SHELLQs quasars we determined in this work. The formation redshift is assumed to be $z = 30$ when the first stars and galaxies were thought to have formed \citep{2004ARA&A..42...79B, 2011ARA&A..49..373B}. The radiation efficiency of $\eta = 0.1$ (i.e. a standard thin accretion disk; \citealt{1976MNRAS.175..613S}) is assumed. We consider two cases. The first we assume is that the SHELLQs quasars had grown at the Eddington limit, and the second is that they had grown with a constant Eddington ratio estimated at their observed redshift. Figure \ref{fig:growth__1} displays the first case; most of the seeds have the mass range of putative Pop-III star remnants ($M_{\rm seed} = 10-100 \ M_{\odot}$) already at $z \sim 20$, in contrast to the cases of previously-studied luminous quasars \citep[e.g.][]{ 2011Natur.474..616M,2015Natur.518..512W,2020ApJ...897L..14Y,2020ApJ...896...23W}. This is primarily because most of our sample have $M_{\rm BH} < 10^{9} M_{\odot}$, and it is possible for Pop-III seeds to create such SMBH masses without growth beyond the Eddington limit. At the same time, some of our sample have masses exceeding the range of the remnant Pop-III stars, in particular those at the highest redshift.
Such objects may be formed from other seed populations such as dense star clusters or direct collapse black holes, or have grown with super-Eddington accretion. 

\begin{figure}[htbp]
\begin{center}
    \includegraphics[height = 5cm,width = 8cm]{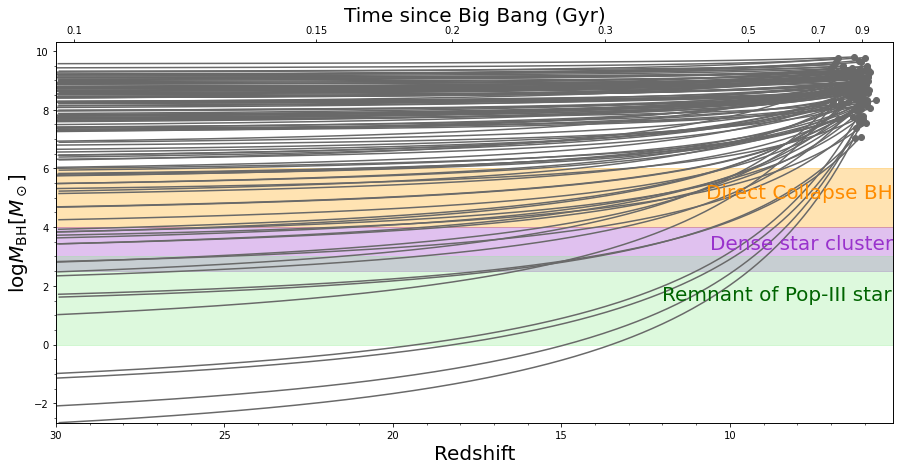}
    \caption{Same as Figure \ref{fig:growth__1}, but for the case of the Eddington ratios being fixed to the observed values.\label{fig:growth__2}}
    \end{center}
\end{figure}

Figure \ref{fig:growth__2} displays the second case, the growth path of our sample with perhaps the more realistic assumption on the Eddington ratios. Most seeds exceed the range allowed for Pop-III remnants, which indicates either that we have to assume other seed populations or that they have grown with higher Eddington ratios than the values observed at $z \sim 6$. Figure \ref{fig:growth__2} may suggest that the distribution of BH mass and the growth history of the high-$z$ low luminosity quasars are divided into the following two phases. The majority of the sample are indicated to have massive seeds even above those predicted by the direct collapse scenario with the assumption of constant observed Eddington ratios. Hence, they are likely to have grown with higher Eddington ratios in the past, and have switched to the less active phase by $z \sim 6$. The others could be in the quiet stage with sub-Eddingon accretion from the seeds to the observed epochs. They may switch to the active phase if they eventually experience major evolutionary events such as galaxy mergers. Sixteen SHELLQs quasars have been observed at the sub-mm wavelength with ALMA, one of which has a companion separated by 15 kpc (Izumi et al. in preparation). This quasar may possibly switch to active mode in a relatively short timescale through a galaxy merger. \\

\subsection{Spectral properties compared with luminosity-matched quasars}

In this section, we compare the spectral properties of our sample with the local luminosity-matched sample, 
\begin{figure*}[th]
\begin{center}
    \includegraphics[height = 5cm,width = 15cm]{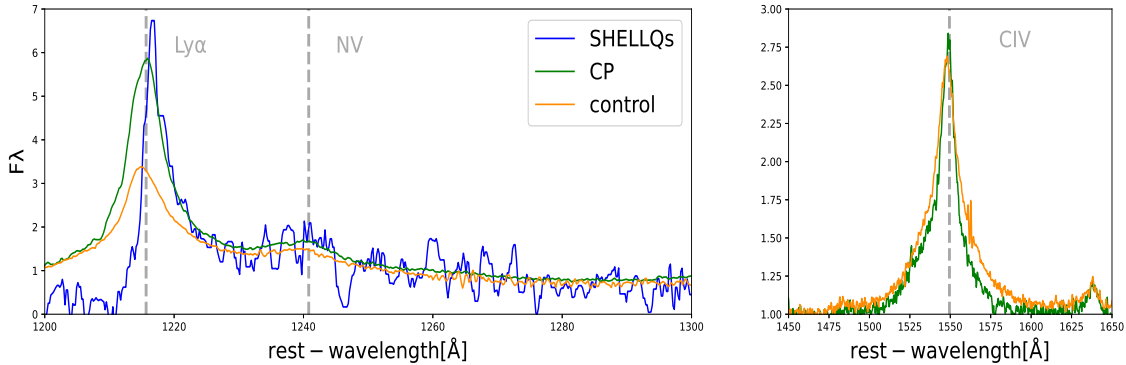}
    \caption{Composite spectra of the SHELLQs (blue), counterpart (green), and control (orange) sample at $\lambda_{\rm rest} =$ 1200 --1300 \AA \ (left) and around C~{\sc iv} (right).  \label{fig:composite_3samples}}
    \end{center}
\end{figure*}

\noindent 
in order to test whether the relation between luminosity and spectral shape has redshift dependence.

Many previous studies have addressed the redshift dependence of quasar spectral properties.
It is generally thought that quasars have little redshift evolution in their spectral properties \citep[e.g.,][]{mortlock2011luminous, shen2019gemini}. \citet{2008MNRAS.390.1413F} and \citet{2008ApJ...680..169S} found that the line widths of the virial BH mass estimators do not depend strongly on luminosity or redshift, especially for luminous quasars in their sample. 
On the other hand, \citet{2010AJ....140..546W} suggested that such results are caused by a wide range of Eddington ratios in the sample used for the analysis. Instead, they found a strong correlation between luminosity and Mg~{\sc ii} FWHM (and thus $M_{\rm BH}$) of the high-$z$ quasars from SDSS and CFHQS, which is presumably due to the fact that most of them are accreting at close to the Eddington limit. W10 also report that the Eddington ratio distribution has a peak at 1.07, which is considerably higher than that of the luminosity-matched sample at $z = 2$ ($\lambda_{\rm Edd}$ peaks at 0.37), suggesting that typical quasars at the higher redshift have higher level of accretion activity.

We now proceed to check the rest-frame UV spectral properties of the SHELLQs sample, which represents the largest sample of $z \geq 6$ quasars at the faint end.
As the first step, we created a low-$z$ control sample matched in continuum luminosity at the rest-frame 1350 \AA \ to our high-$z$ sample. As is well known, quasar luminosity governs many emission line properties, thus it is crucial to match in luminosities. For instance, in high-ionization lines such as C~{\sc iv}, EW decreases with luminosity (i.e., the Baldwin effect, \citealt{1997ASPC..113...80B}), and other line profiles (e.g., velocity shift and asymmetry) also change with luminosity in a systematic manner \citep[e.g.,][]{2002AJ....123.2945R}.  We randomly selected 50 control quasars from the SDSS DR14 quasar catalog within the luminosity difference of $\mathrm{\Delta log L = 0.2}$ for each of the SHELLQs quasars. The control quasars have redshifts $z \sim 1.5 - 5.0$; thus, their spectra cover most of the major rest-frame UV lines from Ly$\alpha$ to Mg~{\sc ii}.  As the second step,
we created the mean composite spectra from each of the SHELLQs sample, counterpart sample, and control sample.

\begin{deluxetable*}{cccccc}
\tablecaption{Spectral measurements from the composites\label{tab:compositelist}}
\tablehead{
\colhead{Sample} & \colhead{FWHM\_Ly$\alpha$} & \colhead{EW\_Ly$\alpha$} & \colhead{FWHM\_C~{\sc iv}} & \colhead{EW\_C~{\sc iv}} & \colhead{LOG\_L1350} \\
 & \colhead{$\mathrm{(km\ s^{-1}}$)} & \colhead{(\AA)} & \colhead{$\mathrm{(km\ s^{-1}}$)} & \colhead{(\AA)} & \colhead{($\mathrm{erg\ s^{-1}}$)}
}
\decimals
\startdata
SHELLQs & 886$\pm$103 & 23$\pm$1.7 &$\cdots$ & $\mathrm{\ }\cdots$ & 45.5\\
counterpart & 1582$\pm$440 & 20$\pm$0.5 & 2294$\pm$655 & 36$\pm$0.4 & 46.0\\
control & 1771$\pm$25 & 11$\pm$0.1 & 3352$\pm$418 & 43$\pm$0.9 &  45.5\\
\enddata
\end{deluxetable*}



Figure \ref{fig:composite_3samples} displays the  composite spectra of the three samples. The Ly$\alpha$ of the SHELLQs and counterpart composite appears to be strong compared to the control, while we see no significant difference in the C~{\sc iv} line between the counterpart and control. We measured EW and FWHM of these lines as follows. We subtracted the continuum fluxes estimated at the both sides of the lines, then fitted the residuals with two or three Gaussian profiles. The SHELLQs composite lacks the blue side of their Ly$\alpha$ line due to the IGM absorption, so we obtained the best-fit model with only the red side of the peak. The associated errors were estimated with 100 mock spectra created via the Markov Chain Monte Carlo method using the observed and propagated noise array. The results of these measurements are shown in Table \ref{tab:compositelist}. FWHM of Ly$\alpha$ is apparently smaller than others. 
On the other hand, Ly$\alpha$ EWs of the SHELLQs and counterpart are twice as high as the control. Since the SHELLQs and the control samples are matched in luminosity, this difference does not reflect the well-known Baldwin effect. However, interestingly, EWs of C~{\sc iv} show the opposite tend; the counterpart has a slightly smaller EW than the control. We suggest that the different behavior of Ly$\alpha$ and C~{\sc iv} may reflect the difference in physical condition of the emitting gas, such as the density and metalicity. The counterpart has smaller line widths than the control in both Ly$\alpha$ and C~{\sc iv} lines, but the difference is within 1sigma uncertainty. 

\begin{figure*}[t]
\begin{center}
    \includegraphics[height = 5cm,width = 15cm]{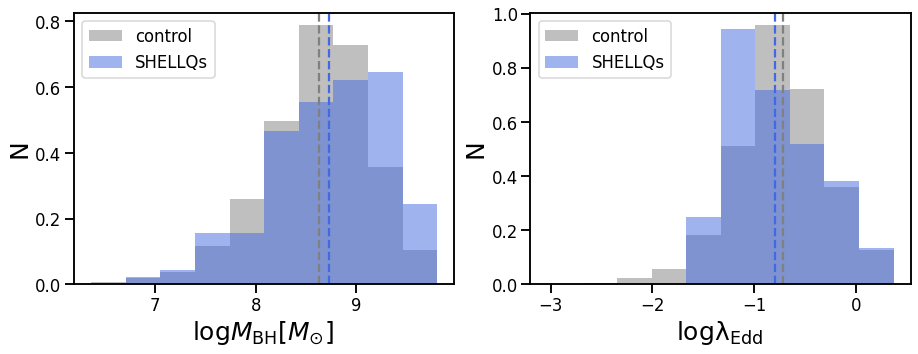}
     \caption{Distributions of BH mass (left) and Eddington ratios (right) of the SHELLQs (blue) and the control (gray) quasars. \label{fig:dist_mbh_redd}}
    \end{center}
\end{figure*}

Figure \ref{fig:dist_mbh_redd} compares the $M_{\rm BH}$ and $\lambda_{Edd}$ distributions of the SHELLQs and the control sample. The p-values of the K-S test in the BH mass and Eddington ratio distribution are $\sim 0.008$ and $\sim 0.0007$, respectively, which may suggest that the basic properties of mass accretion are different between the two samples. The former sample has higher BH masses by 0.1 dex with lower Eddington ratios by 0.1 than the latter. 
At the face value, it may provide a hint of higher SMBH activity in the SHELLQs quasars at $z\sim 6$ than the control sample, despite the fact that the quasar activity generally peaks around $1 < z < 3$.
Some previous studies found more prominent differences between high-$z$ and low-$z$ quasars \citep[e.g.][]{2010AJ....140..546W,shen2019gemini}; the discrepancy from the present work may be attributed to the fact that we are studying unprecedentedly low-luminosity sample at such high redshifts. Having said that, we stress that the present comparisons are subject to strong systematic biases, such as those caused by sample definition, completeness, measurements errors, and assumptions used in the calculations. For robust

\noindent
detection of a 0.1-dex level difference in Eddington ratio distributions for example, one needs to resolve the complexity of those biases and correct for their combined effect, which would be a subject of future projects.

\subsection{Final remarks}

This paper presents a novel method to measure the distributions of BH mass and Eddington ratio among high-$z$ quasars, applicable even to quasars with very low luminosities, such as those established in the SHELLQs program. Our obvious next step is to confirm the estimated properties with direct measurements of a part of the present objects, through future observations with NIR spectrographs either from the ground or using \textit{James Webb Space Telescope}.
Once confirmed with follow-up NIR spectroscopy, the present sample and measurements provide new insight into the early quasar evolution in the reionization epoch.

While we identified candidates of very low-mass SMBHs in our sample, the HSC imaging may have detected more similar objects. The SHELLQs spectroscopy program prioritized candidates with detection in both the HSC $z$ - and $y$-bands, which contain the redshifted Ly$\alpha$ and continuum, respectively.  However, the identified low-mass quasars (e.g., J0859$+$0022) tend to have a very strong and narrow Ly$\alpha$ and faint continuum, and thus are often undetected in the $y$-band. We consider them as the high-$z$ analogs of narrow-line Seyfert 1 found in the local universe. There may be a significant number of similar objects in the remaining HSC candidates awaiting spectroscopy, which may populate the low-mass end of the $M_{\rm BH}$ distribution.

\vspace{5mm}

This research is based on data collected at the Subaru Telescope, which is operated by the National Astronomical
Observatory of Japan. We are honored and grateful for the opportunity of observing the universe from Maunakea,
which has cultural, historical, and natural significance in Hawaii.
The data analysis was in part carried out on the open-use data analysis computer system at the Astronomy Data Center of NAOJ.\\
This work is also based on observations made with the GTC, installed at the Spanish Observatorio del Roque de los
Muchachos of the Insituto de Astrof\'{i}sica de Canarias, on the island of La Palma.\\
Y.M. was supported by the Japan Society for the Promotion of Science (JSPS) KAKENHI grant No. JP17H04830, No. 21H04494, and the Mitsubishi Foundation grant No. 30140.
The HSC collaboration includes the astronomical commu- nities of Japan and Taiwan, as well as Princeton University. The HSC instrumentation and software were developed by the National Astronomical Observatory of Japan (NAOJ), the Kavli Institute for the Physics and Mathematics of the Universe (Kavli IPMU), the University of Tokyo, the High Energy Accelerator Research Organization (KEK), the Academia Sinica Institute for Astronomy and Astrophysics in Taiwan (ASIAA), and Princeton University. Funding was contributed by the FIRST program from the Japanese Cabinet Office; the Ministry of Education, Culture, Sports, Science and Technol- ogy (MEXT); the Japan Society for the Promotion of Science (JSPS); Japan Science and Technology Agency (JST); the Toray Science Foundation; NAOJ; Kavli IPMU; KEK; ASIAA; and Princeton University.
This paper is based on data retrieved from the HSC data archive system, which is operated by Subaru Telescope and Astronomy Data Center (ADC) at NAOJ. Data analysis was in part carried out with the cooperation of Center for Computa- tional Astrophysics (CfCA) at NAOJ.
This paper makes use of software developed for Vera C. Rubin Observatory. We thank the Rubin Observatory for making their code available as free software at http://pipelines.lsst.io/.
The Pan-STARRS1 Surveys (PS1) and the PS1 public science archive have been made possible through contributions by the Institute for Astronomy, the University of Hawaii, the Pan-STARRS Project Office, the Max Planck Society and its participating institutes, the Max Planck Institute for Astron- omy, Heidelberg, and the Max Planck Institute for Extra- terrestrial Physics, Garching, Johns Hopkins University, Durham University, the University of Edinburgh, the Queen’s University Belfast, the Harvard-Smithsonian Center for Astro- physics, the Las Cumbres Observatory Global Telescope Network Incorporated, the National Central University of Taiwan, the Space Telescope Science Institute, the National Aeronautics and Space Administration under grant No. NNX08AR22G issued through the Planetary Science Division of the NASA Science Mission Directorate, the National Science Foundation grant No. AST-1238877, the University of Maryland, Eotvos Lorand University (ELTE), the Los Alamos National Laboratory, and the Gordon and Betty Moore Foundation.


%




\appendix

\section{Spectra of SHELLQs and counterparts}

\begin{figure*}[htbp]
\begin{center}
\includegraphics[height = 5cm, width =10cm]{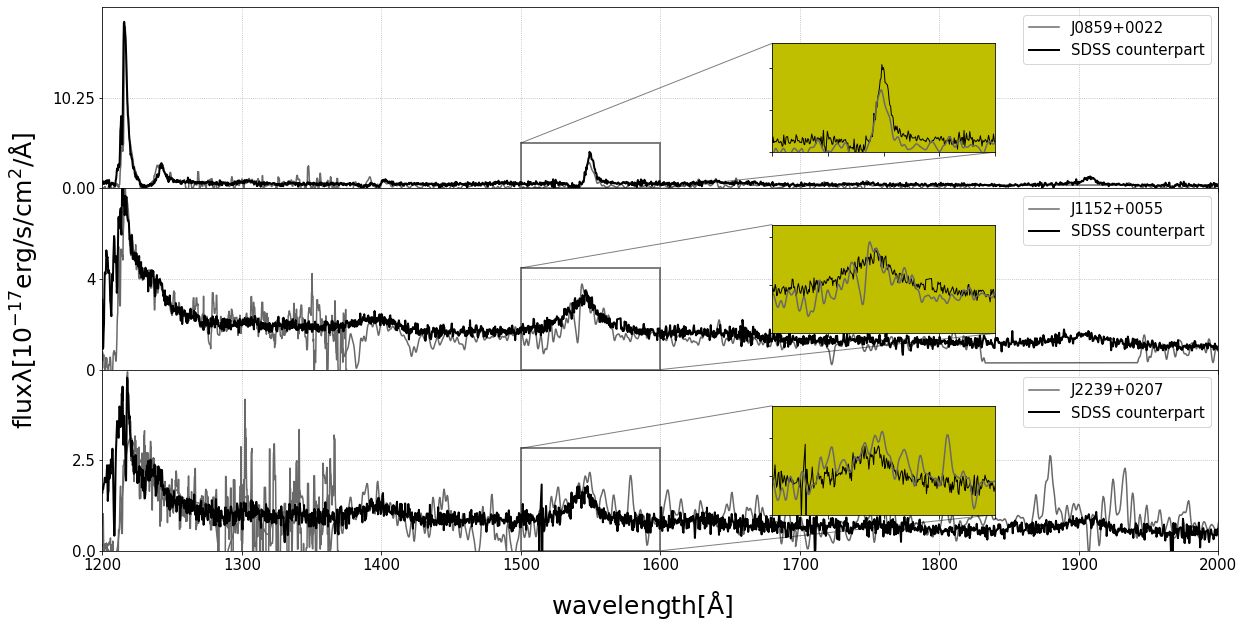}
\caption{Same as Figure \ref{fig:spec_civ_onoue} but for the case in which the SDSS counterparts are obtained with [C ~{\sc ii}] redshift.\label{fig:spec_civ_calbz}}
\end{center}
\end{figure*}


\begin{figure*}[b]
    \begin{tabular}{cc}
      \begin{minipage}[c]{0.8\hsize}
        \centering
        \includegraphics[height = 3.2cm, width = 9.4cm]{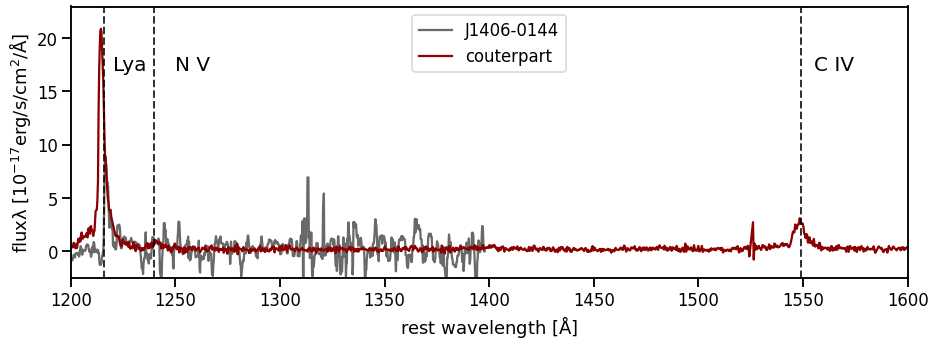}
        \caption{Same as Figure \ref{fig:ex_cp}, but for the specific case of the SHELLQs quasar with super Eddington accretion ($\lambda_{\rm Edd}=1.5$). The Ly$\alpha$ looks very narrow, but its FWHM of 1600 km/s (Matsuoka et al. 2018, ApJS, 237, 5) clearly suggests that it is a type-1 quasar.}
        \label{fig:spec_superedd}
      \end{minipage}\\ 
      \begin{minipage}[c]{0.8\hsize}
        \centering
        \includegraphics[height = 3.2cm, width = 9.4cm]{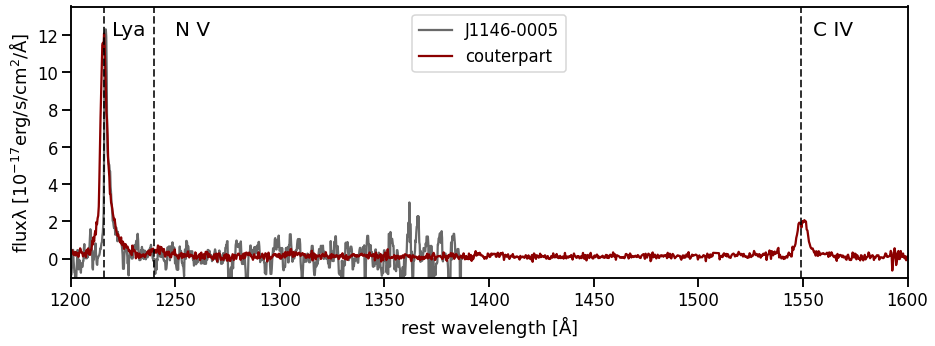}
        \caption{Same as Figure~\ref{fig:ex_cp}, but for the lowest mass SHELLQs quasars with log$M_{\rm BH} \sim 10^{7.0 \pm 0.2} M_{\odot}$ and with a high Eddington ratio.}
        \label{fig:spec_lowmass}
      \end{minipage}
    \end{tabular}
  \end{figure*}

\newpage
{\small
\bibliography{bib.bib}}
\bibliographystyle{aasjournal}

\end{document}